\documentclass[twoside,english,nofootinbib,preprintnumbers,aps,onecolumn,notitlepage]{revtex4-1}
\usepackage{euscript,amssymb}
\usepackage{amsthm}
\usepackage{graphicx}
\usepackage{amsfonts}
\usepackage{amsmath}
\usepackage{amssymb}
\usepackage{fancyhdr}
\usepackage{epsfig}
\usepackage{color}
\usepackage{esint}
\usepackage{soul}
\usepackage{calc}
\usepackage{enumerate}
\usepackage{afterpage}
\usepackage[papersize={8.5in,11in}]{geometry}

\allowdisplaybreaks

\newcommand{\be}{\begin{equation}}
\newcommand{\ee}{\end{equation}}

\begin{document}

\title{Local version of the no-hair theorem}
\author{Denis Dobkowski-Ry{\l}ko}
	\email{Denis.Dobkowski-Rylko@fuw.edu.pl}
	\affiliation{Faculty of Physics, University of Warsaw, ul. Pasteura 5, 02-093 Warsaw, Poland}
\author{Jerzy Lewandowski}
	\email{Jerzy.Lewandowski@fuw.edu.pl}
	\affiliation{Faculty of Physics, University of Warsaw, ul. Pasteura 5, 02-093 Warsaw, Poland}
\author{Tomasz Paw{\l}owski}
	\email{Tomasz.Pawlowski@fuw.edu.pl}
	\affiliation{Center for Theoretical Physics, Polish Academy of Sciences, Al. Lotnik\'ow 32/46, 02-668 Warsaw, Poland.}
	\affiliation{Faculty of Physics, University of Warsaw, ul. Pasteura 5, 02-093 Warsaw, Poland}
\begin{abstract}   
In this paper we study non-extremal isolated horizons embeddable in 4-dimensional spacetimes satisfying the vacuum Einstein equations  
with  cosmological constant. The horizons are assumed to be stationary to the second order. The Weyl tensor 
at the horizon is assumed to be of the Petrov type D. The corresponding   equation on the intrinsic horizon geometry  is 
solved in the axisymmetric case.  The family of the solutions is $2$-dimensional and it is parametrized by the area and 
the angular momentum. The embeddability  in  the Kerr - de Sitter, the Kerr - anti de Sitter 
and the Near extremal Horizon spacetimes obtained by the Horowitz limit from the extremal  Kerr - de Sitter and 
extremal Kerr - anti de Sitter is discussed. This uniqueness of the axisymmetric type D isolated horizons is a generalization 
of an earlier similar result valid in the  cosmological constant free case. 
\end{abstract}

\date{\today}

\pacs{???}

\maketitle
\section{Introduction}

The theory of the non-expanding horizons (NEH) represents the so-called quasi-local approach to describing black holes \cite{ABL1}. Its "Locality" manifests itself via application to that description null surfaces of codimension 1, having the properties of black hole horizons. These null surfaces are assumed to have compact space-like slices of co-dimension 2, which makes the theory only "Quasi" local. That compactness distinguishes the horizons from i.e. planes that are less unusual. The theory is also applicable to spacetimes containing cosmological horizons and to the null boundaries of the conformally compactyfied asymptotically flat spacetimes. Many properties of NEHs are similar to those of black hole spacetimes with many laws/theorems in standard black hole physics having their quasi-local analogs. In particular, there are NEH mechanics (an analog of the black hole "thermodynamics")\cite{ABL2}, the rigidity theorem \cite{LPsymmetric} and the uniqueness theorems \cite{LPkerr, LPextremal}. It is possible to formulate and prove the laws of mechanics (often called the black hole thermodynamics).  Every NEH  is also shear-free and may be called a Killing horizon to the zeroth order.
In the current paper we impose additional assumptions  that make it a Killing horizon to the second order.
We call NEHs that satisfy those assumptions stationary to the second order.  In the non-extremal  case, the spacetime Weyl tensor is determined on a stationary to the second order horizon, via the
$\Lambda$-vacuum Einstein equations and by its intrinsic geometry: a degenerate metric and a consistent torsion free covariant derivative. Hence the Petrov type (normally characterizing the spacetime) can be associated with it.  The intrinsic geometry  is constrained by  all those assumptions, but still,  stationary to the second order horizons have infinitely many degrees of freedom. Therefore, a particularly relevant question is what distinguishes the famous 2-dimensional family of the Kerr, or , more generally, Kerr (anti) de Sitter horizons. A geometric answer to that question in the $\Lambda = 0$ was found in \cite{LPkerr}.  In  that paper we formulated an equation on a complex valued invariant of stationary to the second order horizon geometry that is necessary and sufficient for the Petrov type D of the corresponding Weyl tensor at the horizon. We also found a complete set of axisymmetric solutions to the Petrov type D equation on topologically $S_2\times \mathbb{R}$ horizons. They set a $2$-dimensional family and each of them can be identified either with the inner or the outer  horizon of the non-extremal Kerr metric, or it is the horizon of the near horizon geometry  obtained as the Horowitz-Bardeed limit of the extremal Kerr metric \cite{Horowitz, LPJfol, NHG, LSW}.  Those results are generalized in a series of recent papers \cite{DLP1, LS, DKLS}.  The Petrov type D equation   was derived  in  the case of non-vanishing $\Lambda$ \cite{DLP1}. Its properties and various forms (the covariant, the holomorphic) were studied and discussed.  The  equation was shown to be a necessary integrability condition for the near horizon geometry equation \cite{LPextremal, NHG}.  The emergence of the latter equation in the context of the Petrov type D horizons was shown to be related to  non-twisting of the transversal  principal null direction of the Weyl tensor. The notation introduced in that accompanying work \cite{DLP1} that is useful for the current paper  is outlined in the Appendix A. For stationary to the second order horizons of topology $S\times \mathbb{R}$ where genus of $S$ is bigger than $0$ the Petrov type D equation was  solved  and a complete family of solutions was found  \cite{DKLS}.  In the case of topologically spherical  cross-section $S$                
all the axisymmetric solutions to the  Petrov type D equation with $\Lambda\not=0$ are derived  in the current paper. The family of solutions is $2$-dimensional. The corresponding Petrov type D horizons are embeddable either in a Kerr - (anti) de Sitter spacetime or in a near horizon geometry spacetime. Finally, it turns out, that the assumption of the axial symmetry   can be replaced by assuming that the horizon is bifurcated. Indeed,  a bifurcated horizon stationary to the second order and of the Petrov type D necessarily admits a second, space like symmetry \cite{LS} (See also \cite{R3}\footnote{We kindly ask the reader not to hold us accountable for other  than the axial symmetry in the $\Lambda=0$ case conclusions  stated in \cite{R3}, because that paper has  appeared very recently, well after  the submission of the current work for publication in PRD}).   

Our research on  the Petrov type D stationary to the second order horizons is a part of the  wider context of
detecting geometric horizons \cite{Hay, CM1, CM2, CM3} and distinguishing the Kerr spacetime geometry \cite{Mars1, Mars2}.  Geometric horizons are detected by suitable scalars constructed from the Riemann curvature. In particular, the geometric horizon "detectors" are tuned to exactly to the Petrov type II/D of the Weyl tensor.  To detect the Kerr metric, one needs to use a Killing vector.  A detailed relation between those programs and our current paper is an interesting subject and the work on that is in progress.     

The Petrov type D horizons we consider should not be confused with the Petrov type D spacetimes \cite{PD, Kin}.  The two topics are related but not equivalent.  For example, the Petrov type II Robinson - Trautman vacuum spacetimes contain the Petrov type D horizons  (more precisely, those Robinson-Trautman solutions are of the Petrov Type II except for the points of the horizons where
the Weyl tensor type changes appropriately). A complete understanding of the relation  is an open problem. For example,  one of the results of the current 
paper is  the embedding of axisymmetric Petrov type D horizons in Petrov type D spacetimes.

\subsection{Non-expanding shear free null surfaces} 
Consider  an $n$-dimensional  spacetime  $(M,  g)$ of the signature $-+...+$. 
Suppose $H$ is a non-expanding, shear-free null surface of the co-dimension $1$ embedded in $M$.  Then, the degenerate metric tensor $g_{ab}$ induced in $H$ satisfies
\begin{equation}\label{Llg}
{\cal L}_\ell g_{ab}\ =\ 0 \ = \ell^a g_{ab}
\end{equation}
for every null vector field $\ell$ defined on and tangent to $H$. Moreover, the spacetime covariant derivative $\nabla_\alpha$ 
(metric and torsion free) preserves the tangent bundle of  a non-expanding, shear free null surface $H$  and via the restriction 
endows $H$ with a torsion free covariant derivative $\nabla_a$. The derivative satisfies the pseudo-metricity condition 
\begin{equation}
\nabla_c g_{ab}\ =\ 0,  
\end{equation}
however it is not sufficient to determine $\nabla_a$. 
The pair $(g_{ab}, \nabla_a)$ sets the intrinsic geometry of a null, non-expanding and shear-free surface $H$.  
Moreover, for every null vector field $\ell$ tangent to $H$, there is a 1-form  $\omega^{(\ell)}$ such that 
\begin{equation}\label{omega}
\nabla_a \ell\ =\ \omega^{(\ell)}_a\ell.  
\end{equation}
In particular 
\begin{equation}
\ell^a \nabla_a \ell^b\ =\  \kappa^{(\ell)}\ell^b, \ \  \ \ \kappa^{(\ell)}\ =\ \omega^{(\ell)}_a\ell^a, 
\end{equation}
hence $\kappa^{(\ell)}$ is a self acceleration of $\ell$. 
While the rotation 1-form potential transforms with rescaling $\ell$,
\begin{equation}
 \omega^{(f\ell)} = \omega^{(\ell)} + d{\rm ln}f,
\end{equation}
its external derivative 
\begin{equation}
\Omega =  d\omega^{(f\ell)}
\end{equation}
is a rotation 2-form invariant. 

If the spacetime metric $g$ satisfies the Einstein equations with possibly non-zero  cosmological constant
(or equations of  Einstein's  gravity  coupled to matter fields that satisfy the standard energy inequalities) the self acceleration 
and the 1-form rotation  potential  satisfy the so-called  zeroth low
\cite{ABL2}
\begin{equation}\label{zeroth}
d\kappa^{(\ell)} = {\cal L}_{\ell}\omega^{(\ell)}. 
\end{equation}
We can always choose the vector field $\ell$ to be $\ell_o$ such that  $\nabla_{\ell_o}\ell_o=0$.
Then  
\begin{equation}\label{lo}
0= d\kappa^{(\ell_o)} = {\cal L}_{\ell_o}\omega^{(\ell_o)}. 
\end{equation}
Now using such  $\omega^{(\ell_o)}$ we calculate $\Omega$. Then,  the eqs. (\ref{lo}) imply 
\begin{equation}\label{LlOmega}
{\cal L}_{\ell}\Omega = 0 = \ell^a \Omega_{ab}.
\end{equation}
Given a null field $\ell$ one can introduce a coordinate $v$ compatible with $\ell$, namely such that $\ell_a v_{,a} = 1$. That coordinate in turn defines the covector field $n_a := -v_{,a}$.

\subsection{Isolated null surfaces} 
A null, non-expanding shear-free surface  $H$ admitting a null vector $\ell$ such  that
\begin{equation}\label{LlD}
[{\cal L}_\ell, \nabla_{a}]\ =\ 0
\end{equation}
is called isolated. We are also assuming throughout this paper, that 
$$\ell\not=0$$
at every point of $H$.  
Notice that while the condition (\ref{Llg})  is preserved by every transformation $\ell \mapsto f\ell$ with arbitrary function
 $f$ defined on $H$, the condition (\ref{LlD})   generically is invariant only for 
 $$f= f_0={\rm const}\not= 0.$$
 In the case of isolated null surface the equalities (\ref{omega}) and (\ref{LlD}) imply
 \begin{equation}\label{Llomega}
{\cal L}_{\ell}\omega^{(\ell)} = 0,
\end{equation}
hence eq. (\ref{zeroth}) takes the following form
\begin{equation}\label{dkappa}
d\kappa^{(\ell)} = 0,
\end{equation}
meaning that the self acceleration  $\kappa^{(\ell)}$ is constant along (every connected  part) the null surface $H$. 
The rescaling of a null symmetry generator $\ell$ by a constant results in rescaling  the self acceleration 
\begin{equation}
\kappa^{(f_0\ell)} \ =\ f_0\kappa^{(\ell)} . 
\end{equation}
Therefore,  there are essentially two cases of an isolated null surface: either non-extremal when
\begin{equation}
\kappa^{(\ell)}\ \not= \ 0
\end{equation}
or extremal,  whenever
\begin{equation}
\kappa^{(\ell)}\ = \ 0. 
\end{equation}
There exist exceptional  cases, though,  of isolated null surfaces that admit two dimensional null symmetry group,
and are non-extremal with respect to one generator $\ell$ and extremal with respect to another symmetry
$f\ell$,  where $\nabla_af\not=0$  \cite{ABL1}.  Because of those special cases, while considering  isolated surfaces we 
also indicate the vector $\ell$  (up to constant rescaling)  that generates a null symmetry (\ref{LlD}).   

In case the isolated null surface is contained in an Einstein spacetime, that is when
 \begin{equation}\label{E}
{}^{(n)}R_{\alpha\beta}-\frac{1}{2}{}^{(n)}R g_{\alpha\beta} + \Lambda g_{\alpha\beta}\ =\ 0
\end{equation}       
the equations constrain the isolated horizon intrinsic geometry $(\ell, g_{ab}, \nabla_a)$ to the extend,
that in a non-extremal case   the internal connection  $\nabla_a$ is fully determined by:  $g_{ab}$, $\ell$, and  
$\omega^{(\ell)}$. That data is free modulo the constraints (\ref{Llomega}, \ref{dkappa}). 
  
\section{Isolated horizons}
An isolated horizon is any isolated null surface $(H, \ell)$ diffeomorphic to $\mathbb{R}\times S$, where $S$ is  a compact manifold,  and such that every slice of $H$  corresponding to $S$ is spacelike. In that case, it follows from eqs. (\ref{Llg}) and (\ref{LlOmega}) that  the degenerate metric tensor $g_{ab}$ and the rotation 2-form invariant $\Omega_{ab}$ are determined by 
 a (Riemannian)  metric tensor $g_{AB}$ and an exact 2-form $\Omega_{AB}$ defined on the quotient manifold $S$ via a pullback of the map
 $$\Pi : H \mapsto S ,$$
 \begin{align}\label{hatqOmega} g_{ab}\ =\ \Pi^*{}_{ab}{}^{AB}g_{AB}, \ \ \ \ \ \ \ \ \ \ \  \ \Omega_{ab}\ =\ \Pi^*{}_{ab}{}^{AB}\Omega_{AB} \end{align}
 
In our paper, henceforth we will be assuming that the dimension of the  spacetime   $(M, g)$ is 
$$n=4,$$ 
the spacetime metric $g_{\alpha\beta}$ satisfies the Einstein equations (\ref{E}) with arbitrary cosmological constant $\Lambda$, and  $S$ is topologically the 2-dimensional sphere $S_2$, $$ S =\ S_2.$$ According to (\ref{hatqOmega}), the 2-sphere is endowed with a metric tensor $g_{AB}$, it's covariant derivative $\nabla_{A}$, and a 2-form $\Omega_{AB}$ that can be written in terms of a scalar invariant and area 2-form ${}^{{}^{(2)}}\epsilon_{AB}$ 
 \begin{equation}
 \Omega_{AB} =: \mathcal{O} \ {}^{{}^{(2)}}\epsilon_{AB}.
 \end{equation}
 From the freely given $g_{AB}$ and $\omega_A$, both, $K$ and $\mathcal{O}$ can be derived respectively. The rotation scalar invariant $\mathcal{O}$ can be expressed by an a priori unconstrained function
$ U:S_2\rightarrow \mathbb{R},$ as
 \begin{equation}
 \mathcal{O}= - \nabla^A \nabla_A U.
 \end{equation}  
 The function $U$ is often reffered to as rotation potential. The Gaussian curvature $K$ of the metric tensor $g_{AB}$ will be another scalar invariant used to characterize the isolated horizons $H$. 
  
\section{The type D equation}
It is convenient now,  to express the components of IH intrinsic geometry in terms of coefficients in Newman-Penrose null tetrad $(e_1,...,e_4)$ defined as follows:
\begin{itemize}
\item $e_1^\mu := m^\mu$, $e_2^\mu := \bar m^\mu$, $e_3^\mu := n^\mu$, $e_4^\mu :=\ell^\mu$.
\item $e_1, e_2 := \bar e_1$ are tangent to constancy surfaces of a coordinate compatible with $\ell$ and Lie-dragged along null curves of the horizon.
\item The only non-vanishing products of $e_I$ with respect to spacetime metric are the following ones:
\begin{align}
g_{\mu \nu} e_1^\mu e_2^\nu = 1, && g_{\mu \nu} e_3^\mu e_4^\nu = -1.
\end{align}
\end{itemize}

In the defined frame the horizon intrinsic geometry components are determined by complex connection coefficients in the following way:
\begin{itemize}
\item  $\nabla_a \ell^b  = \omega_a^{(\ell)}  \ell^b$, where  $\omega_a^{(\ell)} =  (\alpha + \bar \beta)m_a + (\bar \alpha +\beta)\bar m_a - (\epsilon +\bar \epsilon)n_a$;
 \item $\bar m^b  \nabla_an_b = \lambda m_A + \mu \bar m_a - \pi n_a$;
 \item $m^b  \nabla_a \bar m_b = -(\alpha - \bar \beta)m_a + (\bar \alpha - \beta) \bar m_a + (\epsilon - \bar \epsilon) n_a  = - \bar m^b  \nabla_a m_b$,
\end{itemize}
 The 1-form $\omega^{(\ell)}_A := (\alpha + \bar \beta)m_A + (\bar \alpha +\beta)\bar m_A$ can be expressed via Hodge decomposition in terms of rotation potential $U$ and a function $B$ encoding respectively its exact and coexact parts:
\begin{align}\label{eq:omega}
\omega^{(\ell)}_A = d \ln B_A + \star d U_A.
\end{align}
The coexact part is always invariant  whereas the exact part is a pure gauge provided that $\kappa^{(\ell)} \neq  0$. In particular $ d \ln B$ can be set to zero by an appropriate choice of $v$. This gauge condition will be used from now on. 

In Newman-Penrose formalism the Weyl tensor is described in terms of five complex coefficients $\Psi_0$, ..., $\Psi_4$,
\begin{equation}
\Psi_0\ =\ C_{4141}, \ \ \ \ \ \Psi_1\ =\  C_{4341}\ \ \ \ \ \Psi_2\ =\ C_{4123}, \ \ \ \ \ 
\Psi_3\ =\ C_{3432}, \ \ \ \ \ \Psi_4\ =\   C_{3232} .
\end{equation}
On the horizon $\Psi_0$ and $\Psi_1$ vanish identically, whereas $\Psi_2$ is a scalar invariant:
\begin{equation} \label{eq:psi21}
\Psi_{2} = \bar\delta \beta - \delta \alpha + (\mu\rho - \lambda\sigma) + \alpha\bar\alpha + \beta\bar\beta - 2\alpha\beta + \gamma(\rho-\bar\rho) + \epsilon(\mu - \bar\mu)   + \Lambda/6,
\end{equation} 
where $\rho$ and $\sigma$ are all equal to zero on a vacuum non-expanding null surface. From now on we will denote 
$\Lambda/6$ as $\Lambda'$. In the tangent frame $(m, \bar m, n, l)$ introduced above additional identities hold (see Appendix B):  
\begin{equation}
\mu=\bar\mu, \\
\end{equation}
\begin{equation}
\pi=\alpha + \bar\beta, \\
\end{equation}
\begin{equation}
{\rm Im}(\epsilon) = 0,
\end{equation}
which have been derived in \cite{LPnei}. The surface gravity $\kappa$ can be expressed in terms of $\epsilon$ and its nonzero value can be arbitrarily fixed:
\begin{equation}
2\text{Re}(\epsilon)= \kappa = \text{const}.
\end{equation}
Eq. (\ref{eq:psi21}) may be then simplified and expressed using the Gaussian curvature $K$ and the scalar invariant $\mathcal{O}$:
\begin{equation} \label{eq:psi22}
\Psi_{2} = - \frac{1}{2} (K + i\mathcal{O}) + \Lambda',
\end{equation}
where:
\begin{align}
K &:= \delta(\alpha - \bar\beta) +  \bar\delta(\bar\alpha - \beta) - 2(\alpha - \bar\beta)(\bar\alpha - \beta), \\
\mathcal{O} &:= - \bigg[ \bar\delta\delta + \delta\bar\delta + (\beta - \bar\alpha)\bar\delta + (\bar \beta -\alpha)\delta \bigg] U.
\end{align}
Also the component $\Psi_3$ at $H$ can be expressed in terms of $m_A$ and $\alpha +\bar \beta$. The component
$\Psi_4$ can be arbitrarily set at a cross-section of $H$ and evolved according to the equations implied by the Bianchi
identities. However, we are additionally assuming that $\Psi_4$ is constant along the null geodesics of $H$
\begin{equation}\label{stationarity}
\ell^a\partial_a \Psi_4=0.
\end{equation}
We refer to that assumption as the stationarity to the second order \cite{DLP1}. It means that there is an extension $t$ of the vector field $\ell$ to a neighborhood of $H$ such that
\begin{equation}\label{t}
 {\cal L}_t g_{\mu\nu}|_{H}\ =\ [{\cal L}_t, \nabla_\mu]|_{H}\ =\ 0\ ={\cal L}_t R_{\mu\nu\alpha\beta}|_{H}.
\end{equation}
The existence of an extension $t$ that satisfies the first two equalities is already ensured by (\ref{Llg}) and (\ref{LlD}). Eq. (\ref{stationarity}) is equivalent to  the third equality.
Then, due to the non-extremality of $H$ and $\ell$, the component  $\Psi_4$ is determined at $H$ by the data $m_A$ and 
$\alpha+\bar{\beta}$ as well.  

The condition that Weyl tensor is of the Petrov type $D$ on $H$ is \cite{DLP1}:
\begin{equation} \label{eq:typeD1}
2\Psi_{3}^2 = 3\Psi_{2}\Psi_{4}.
\end{equation}
Expressing $\Psi_{3}$ and $\Psi_{4}$ in terms of  $m_A$ and $\alpha+\bar{\beta}$ and substituting into eq. (\ref{eq:typeD1}) yields the  type D equation \cite{LPkerr, DLP1}
\begin{equation}\label{eq:typeD}
0= (\bar\delta + \alpha -\bar  \beta) \bar\delta\Psi_2^{-\frac{1}{3}},
\end{equation}
where 
$$\delta := e_1^a\partial_a=m^A\partial_A.$$
 \section{Coordinates adapted to axial symmetry}\label{coordinates}
In the presence of an axial symmetry the coordinate $v$ compatible with $\ell$ can be chosen in a way that the orbits of axial symmetry lie entirely on the constancy surfaces of $v$. This condition is compatible with the requirement that the exact part of rotation 1-form vanishes (both can be satisfied at the same time). Moreover, one can choose the frame vectors $e_{1,2}$ such that  are preserved by axial symmetry. 

We will solve eq. (\ref{eq:typeD}) assuming axial symmetry, but first we need to choose a coordinate system. In this case, particularly useful to work with, is the so-called $system$ $adapted$ $to$ $axial$ $symmetry$ $(x,\phi)$, which is constructed in the following way. We start with a 2-sphere metric and a conformal factor $\Sigma$ (independent of $\varphi$ because of the symmetry):
 \begin{equation}\label{eq:2Smetric}
 g_{AB}dx^Adx^B= \Sigma^2(\theta) (d\theta^2 + \sin^2\theta d\varphi^2).
 \end{equation}
 The area element of the above metric reads:
 \begin{equation}
{}^{{}^{(2)}}\epsilon = \Sigma^2\sin\theta d\theta \wedge d\varphi.
\end{equation}
One could consider the following transformation:
\begin{equation}
dx = \frac{\Sigma^2 \sin\theta}{R^2} d\theta,
\end{equation}
where $R^2$ is defined to be the area radius satisfying: 
$$A= 4\pi R^2.$$ 
We now introduce the function 
$$P^2=\frac{\Sigma^2 \sin^2{\theta}}{R^2},$$ which we will refer to as the frame coefficient. Calculating the area of the transformed metric $g_{AB}$ yields:
\begin{equation}
A = \int {}^{{}^{(2)}}\epsilon = R^2 (x_1 - x_0) 2\pi.
\end{equation}
Since $x$ has been defined up to an additive constant, by setting $x_1=1$ from the above equation we obtain that $x_0=-1$. The coordinate $\varphi$ is such that the (normalized) infinitesimal axial symmetry equals $\partial_\varphi$ and the curves $\varphi=\text{const}$ are orthogonal to the infinitesimal symmetry. 
The metric tensor $g_{AB}$ reads: 
\begin{equation}\label{gAB}
g_{AB}dx^Adx^B = R^2\bigg(\frac{1}{P(x)^2} dx^2 + P(x)^2 d\varphi ^2\bigg),
\end{equation}
whereas the frame vector $e_1$ and its dual have the form:
\begin{align}
e_1= m^A \partial_A = \frac{1}{ R \sqrt{2}}\left(P(x)\partial_x 
                                +i\frac{1}{P(x)}\partial_\varphi\right), && e^1=\bar m_A dx^A=\frac{R}{\sqrt{2}}\left( \frac{1}{P(x)} dx - iP(x) d\varphi \right).       
\end{align}         
The conditions for the regularity of the metric (lack of conical singularity) are as follows:
\begin{align} \label{eq:boundary}
P(x=\pm1)=0, \ \ \ \ \ \ \  \lim_{x\to \pm1} \partial_x P^2 = \mp 2,
\end{align}
where the values $x=\pm 1$ correspond to the poles ($\theta=\pi,0$). Furthermore an analogous assumption of the lack of conical singularity of $U$ imposes the following constraint on the rotation potential:

\begin{align}\label{eq:conditionU}
 \lim_{x\to \pm1} P\partial_x U \ =0 .
\end{align}

Now in this system we can calculate the connection component $\alpha - \bar \beta$, Gaussian curvature $K$ and Laplacian $\nabla^A \nabla_A$
acting on scalar functions: 

\begin{align}
\alpha - \bar \beta =  \bar m^A \delta \bar m_A - m^A \bar \delta \bar m_A &  = - \frac{1}{R\sqrt{2}} \partial_x P \\
K = \delta (\alpha - \bar \beta) + \bar \delta (\bar \alpha - \beta)  - 2 (\alpha - \bar \beta)(\bar \alpha - \beta)   &= -\frac{1}{2 R^2}\partial_{x}^{2}P^2 \\
\nabla^A \nabla_A = \bar\delta\delta +  \delta\bar\delta +  ( \beta - \bar\alpha)\bar\delta + (\bar \beta -\alpha)\delta &= \frac{1}{R^2}  \partial_x(P^2 \partial_x).
\end{align}
Applying these to (\ref{eq:typeD}) we obtain a complex ordinary differential equation for $\Psi_2$:
\begin{equation}\label{eq:ordinary}
\begin{split}
0&=(\frac{1}{R\sqrt{2}}P\partial_x - \frac{1}{R\sqrt{2}}\partial_x P)\frac{1}{R\sqrt{2}}P\partial_x \Psi_2 ^{-\frac{1}{3}} \\
  &=\frac{P^2}{2R^2}\partial_x^2\Psi_2^{-\frac{1}{3}}.
\end{split}
\end{equation}
Integrating equation (\ref{eq:ordinary}) yields:
\begin{equation}\label{eq:psi23}
\Psi_{2}=(c_{1} x + c_{2})^{-3},
\end{equation}
where $c_{1}$, $c_{2}$ are complex constants. Substituting this to (\ref{eq:psi22}) we obtain a second order equation for the frame coefficient $P$ and the potential $U$:
\begin{equation} \label{eq:pu}
\frac{4R^2}{(c_{1} x + c_{2})^{3}}= \partial_{x}^{2}P^2 + 2i\partial_x(P^2\partial_x U) + 4R^2 \Lambda'.
\end{equation}

A very geometric relation between the symmetry generator
$$\Phi=\partial_\varphi$$
and $\Psi_2$ is another consequence of (\ref{eq:psi23}), namely
 \begin{equation}
d(\Psi_2^{-\frac{1}{3}}) = - \frac{c_1}{R^2} \Phi  \mathrel{\lrcorner}  {}^{{}^{(2)}}\epsilon.
\end{equation}
It is known in the $\Lambda=0$ case \cite{LPkerr} and apparently,  it generalizes to $\Lambda\not=0$.

\section{General solution to the type D equation on a 2-sphere for the axisymmetric Isolated Horizon} 
In the following subsections we find the solutions to type D constraint on a 2-sphere using coordinates adapted to the axial symmetry introduced in (\ref{coordinates}). We consider a special case when the imaginary constant $c_1$ vanishes and a generic one when both constants $c_1$ and $c_2$ take arbitrary nonzero values.

\subsection{The non-rotating case}
We first consider the case  
$$c_1=0.$$
 Consequently, eq. (\ref{eq:pu}) reads:
\begin{equation}
\frac{4R^2}{ c_{2}^{3}}= \partial_{x}^{2}P^2 + 2i\partial_x(P^2\partial_x U) + 4R^2 \Lambda'.
\end{equation}
Integrating it once and applying conditions (\ref{eq:boundary}, \ref{eq:conditionU}) we obtain a relation between the area radius $R^2$ and $c_2$:
\begin{equation}\label{rs}
R^2=\frac{1}{2} \frac{-c_2^3}{1-c_2^3 \Lambda'}.
\end{equation}
Since $R^2$ has to be positive, the following must hold:
\begin{align}
c_2^3 &\in \bigg(-\infty,0 \bigg) \vee  \bigg(\frac{1}{\Lambda'}, \infty  \bigg), \ \ \ \ \ \ \ \  \text{for $\Lambda' > 0$}; \\ 
c_2^3 &\in \bigg(- \infty, \frac{1}{\Lambda '} \bigg), \ \ \ \ \ \ \ \ \ \ \ \ \ \ \ \ \ \ \ \ \ \  \text{for $\Lambda' < 0$}.
\end{align}

Furthermore, we substitute the expression for $R^2$ into eq. (\ref{eq:pu}) to find that:
\begin{equation}
-2 = \partial_x^2 P^2 + 2i \partial_x \big ( P^2 \partial_x U \big ),
\end{equation}
and integrate it to calculate the integration constant using boundary conditions for the second time:
\begin{equation} \label{eq:sds}
-2x = \partial_x P^2 + 2i P^2 \partial_x U.
\end{equation}
Now we integrate a real part of the above equation to obtain an expression for $P^2$:
\begin{equation}\label{ps}
P^2 = 1 - x^2.
\end{equation}
Note that substituting $P^2$  in eq. (\ref{eq:sds}) yields:
\begin{equation}
0=(1-x^2) \partial_x U,
\end{equation}
and since it has to be satisfied for all $x \in [-1,1]$ it follows that:
\begin{equation}\label{uconst}
U= {\rm const}.
\end{equation}
In summary, the general solution with $c_1=0$ is given by the function $P^2:[-1,1]\rightarrow \mathbb{R}$ 
 (\ref{ps}),  the function $U$ (\ref{uconst}), and arbitrary constant 
$$R>0.$$  
They give rise to the metric tensor  $g_{AB}$ (\ref{gAB}), and the rotation $1$-form potential
$$\omega^{(\ell)}_A=0.$$ 
 We find that Gaussian curvature of $g_{AB}$ is constant:
\begin{equation}
K = - \frac{1}{2 R^2} \partial_x^2 P^2 = \frac{2}{c_2^3} \Big(c_2^3 \Lambda' - 1 \Big),
\end{equation}
which means tha in this case $g_{AB}$ is a round sphere metric of the radius $R$ (knowing that $A=4\pi R^2$). 
 
Therefore, there exists a coordinate $\theta(x)$ such that our metric tensor on $S$  
takes the canonical form, namely
\begin{equation}
R^2(\frac{1}{P^2} dx^2 + P^2 d\varphi ^2) = R^2(d\theta^2 + \sin^2\theta d\varphi^2).
\end{equation}
Comparing the coefficients and using the fact that in both cases the infinitesimal axial symmetry is suitably normalized  we obtain a relation between  $\theta$ and $x$:
\begin{equation}
1-x^2 = \sin^2 \theta.
\end{equation}
The above equation can be used to find the  function $x(\theta)$:
\begin{equation}
\begin{split}
\frac{R^2}{1-x^2} dx^2 &= R^2 d\theta^2 \\
dx^2 &= \sin^2 \theta d\theta^2 \\
x &=  - \cos \theta .
\end{split}
\end{equation}
Since conversely,  every spherically symmetric solution to the type D equation has this form, the  conclusion is, that the family of spherically symmetric 
type D isolated horizons that satisfy  vacuum Einstein equations with cosmological constant $\Lambda$ is $1$ dimensional and can be parametrized 
by the radius $R$.  

Nextly, we shall proceed to discuss the issue of the embeddability of the $1$-dimensional family of spherically symmetric  non-extremal isolated horizons of the type D in generalized Schwarzschield - (anti) de Sitter spacetime, defined by the static spherically symmetric spacetime metric tensors  satisfying the vacuum Einstein equations with the cosmological constant $\Lambda$: 
\begin{equation}\label{SdS}
ds^2 = - \bigg( 1-\frac{2M}{r} - \frac{\Lambda}{3} r^2  \bigg) dt^2 + \frac{dr^2}{1- \frac{2M}{r} - \frac{\Lambda}{3} r^2} + r^2(d\theta^2 + \sin^2\theta d\phi^2).
\end{equation} 
Every spacetime in this family   is of the Petrov type D. Its extension contains Killing horizons, one for each  root $R$ of the equation
\begin{equation}\label{horR} 1-\frac{2M}{R} - \frac{\Lambda}{3} R^2 =0.\end{equation}
Each of those horizons, as long as it is non-extremal, is one of our  type D horizons.  For every value of the radius parameter $R$ of our solutions,  the
matching  parameter $M$ in (\ref{SdS}) can be found such that $R$ is a root  of (\ref{horR}). The result is 
\begin{equation}
M = \frac{R}{2} \bigg(1- \frac{\Lambda}{3} R^2 \bigg). 
\end{equation}
For 
$$\Lambda>0, \ \ \ \ \ \ \ \ {\rm and}\ \ \ \ \ \ \ R=\sqrt{\frac{3}{\Lambda}}  $$
this becomes the de Sitter metric. There is also a non-trivial possibility of 
$$\Lambda>0, \ \ \ \ \ \ \ \ {\rm and}\ \ \ \ \ \ \ R>\sqrt{\frac{3}{\Lambda}}  $$
that yields 
$$M<0,$$
and a non-physical metric that still contains one of our type D horizons provided it is non-extremal. 

The horizon is non-extremal as long as for every function $f$ the tensor
\begin{align} \label{extrem}
\mu' (m_A\bar{m}_B &+ m_B\bar{m}_A) +\lambda' m_A{m}_B + \bar{\lambda}' \bar{m}_A\bar{m}_{B}  \  \nonumber \\&=\ \frac{1}{\kappa^{(\ell)}} \bigg( \nabla _{(A} (\omega_{B}+f_{,A}) + (\omega_A+f_{,A}) (\omega_B+f_{,A}) - \frac{1}{2} K q_{AB} + \frac{1}{2} \Lambda q_{AB} \bigg)
\end{align}
does not vanish, and is extremal otherwise  \cite{ABL1}.  In the spherically symmetric horizon intrinsic geometry  case,  
it is enough to consider $f$ such that    $\omega_A + f_{,A}$ is zero. Upon that assumption we can  calculate the  right hand side easily, namely:
\begin{equation}
\begin{split}
\nabla _{(A} (\omega_{B}+f_{,A}) + (\omega_A+f_{,A}) (\omega_B+f_{,A}) - \frac{1}{2} K q_{AB} + \frac{1}{2} \Lambda q_{AB} 
=\frac{1}{2 }   \bigg( -\frac{1}{R^2} + \Lambda \bigg) q_{AB}.
\end{split}
\end{equation}
Hence, the Killing horizon in the spacetime (\ref{SdS}) is non-extremal if and only if
$$ R^2 \not= \frac{1}{\Lambda}. $$ 
However we still  have a non-extremal solution to the type D equation also for 
$$c_2^3= - \frac{3}{\Lambda}$$
that is 
 $$R^2= \frac{1}{\Lambda}. $$ 
It is embeddable in the near  horizon geometry  spacetime  obtained by the near extremal horizon  limit from
(\ref{SdS}) \cite{NHG}. In that case, the horizon is both, extremal and non-extremal \cite{ABL1,LSW}.  

\subsection{The rotating case}
Analogously, we will now examine the case of  
$$c_1\not=0.$$  
Integrating eq. (\ref{eq:pu}) once and applying conditions (\ref{eq:boundary}, \ref{eq:conditionU}) yields:
\begin{equation}
\begin{split}
\frac{-2R^2}{c_1 (c_1 x+c_2)2} \Biggr|_{-1}^{1}&= \Bigg [\partial_xP^2  +2iP^2 \partial_x U  + 4R^2 \Lambda x \Bigg ] _{-1}^{1}\\
\frac{-2R^2}{c_1(c_1+c_2)^2}+ \frac{2R^2}{c_1(-c_1+c_2)^2} &= -4 + 8R^2 \Lambda,
\end{split}
\end{equation}
and therefore, it gives an expression for $R^2$ in terms of complex coefficients $c_1$ and $c_2$: 
\begin{equation} \label{eq:r}
R^2 = \frac{(c_{2}^2 - c_{1}^2)^2}{-2c_{2} + 2\Lambda'(c_{2}^2 - c_{1}^2)^2} = \frac{1}{2}\frac{\gamma}{\Lambda'\gamma - 1},
\end{equation}
where we have introduced parameter $\gamma=\frac{(c_{2}^{2}-c_{1}^{2})^2}{c_{2}}$ and assumed that $c_2$ does not vanish\footnote{in case of $c_2=0$ the geometry is not well-defined at $x=0$, therefore we will exclude it from our considerations.}. Globality conditions fix the additive constant freedom:

\begin{equation} \label{eq:pu2}
\begin{split}
\partial_x P^2 + 2iP^2 \partial_x U  = &- \frac{(c_{2}^{2} - c_{1}^{2})^2}{c_{1}(c_{1} x+c_{2})^2[(c_{2}^2 - c_{1}^2)^2\Lambda - c_{2}]} \\  &- \frac{2\Lambda'(c_{2}^{2} - c_{1}^{2})^2 x}{(c_{2}^{2} - c_{1}^{2})^2\Lambda' - c_{2}} \\  &+ \frac{c_{2}^2+c_{1}^2}{c_{1}[(c_{2}^{2} - c_{1}^{2})^2\Lambda'-c_{2}]}.
\end{split}
\end{equation}
Now we can integrate the real part of eq. (\ref{eq:pu2}) to get an expression for $P^2$. Applying condition (\ref{eq:boundary}) we obtain:

\begin{equation}\label{eq:constraint2}
{\rm Re}\bigg[\frac{4c_{1}}{(c_{2}^2-c_{1}^2)^2\Lambda' - c_{2}}\bigg] = 0.
\end{equation}
Since radius $R$ is real, from (\ref{eq:r}) we get that:
\begin{equation} \label{eq:constraint3}
\begin{split}
0  &={\rm Im}\bigg[\frac{(c_{2}^2 - c_{1}^2)^2}{-2c_{2} + 2\Lambda'(c_{2}^2 - c_{1}^2)^2}\bigg] \\ 
&= {\rm Im}\bigg[\frac{c_{2}}{-c_{2} + \Lambda'(c_{2}^2 - c_{1}^2)^2}\bigg].
\end{split}
\end{equation}
This simplifies eq. (\ref{eq:pu2}), which now reads:
\begin{equation}\label{eq:pu3}
\begin{split}
\partial_x P^2 + 2iP^2 \partial_x U  =  -2x + \frac{1}{1-\gamma \Lambda'} \frac{1}{\zeta} \bigg[ \frac{(1-\zeta^2)^2}{(x+\zeta)^2} -\zeta^2-1+2\zeta x \bigg],
\end{split}
\end{equation}
where we have introduced a purely imaginary parameter $\zeta = \frac{c_{2}}{c_{1}}$. Integrating the real part of the above equation and using $\eta=-i\zeta$ we obtain the expression for the frame coefficient $P^2$ of the following form:  

\begin{equation}\label{eq:p2}
P^2= 1-x^2 + \frac{1}{1-\gamma\Lambda'} \frac{(x^2-1)^2}{x^2+\eta^2}.
\end{equation}
Pluggin the expression for $P^2$ into eq. (\ref{eq:pu2}) we find:
\begin{equation}\label{eq:U}
\partial_x U=\frac{1}{2\eta} \frac{3\eta^4 - x^2 +\eta^2(x^2+1)}{\Big( x^2+\eta^2 \Big)\Big(\eta^2+1-\gamma\Lambda'(x^2+\eta^2)\Big)},
\end{equation}
Condition (\ref{eq:conditionU}) is satisfied since:
\begin{align}
P\mid_{x=\pm1} \ &= 0  , \\ 
\partial_x U\mid_{x=\pm1} \ &=\frac{1}{2\eta} \frac{3\eta^4 +2\eta^2-1}{{(1-\gamma\Lambda')( 1+\eta^2 )}^2} , \\ 
 1- \gamma \Lambda ' &\neq 0, \\ 
\eta &\neq 0.
\end{align}
Now, once we have obtained the expression for $\partial_x U$ we use eq. (\ref{eq:omega}) to find the rotation 1-form $\omega^{(\ell)}$:
\begin{align}\label{omega'}
{\omega^{(\ell)}_\varphi}&= \star dU_\varphi= {{}^{(2)}\epsilon_{\varphi x}} g^{xx} dU_{x} =- {P^2} \partial_x U = \frac{(1-x^2)\big[ 3\eta^4+\eta^2 +x^2(\eta^2-1) \big]}{2\eta (\gamma\Lambda'-1)(x^2+\eta^2)^2}.
\end{align}
To conclude, the general solution with the non-vanishing constant $c_1$ is given by the function $P^2:[-1,1]\rightarrow \mathbb{R}$ (\ref{eq:p2}), arbitrary $R^2$ (\ref{eq:r}) and the rotation 1-form ${\omega^{(\ell)}_A}$ (\ref{omega'}), which can all be expressed in terms of parameters $\eta$ and $\gamma$.

\subsection{Parameters' constraints}
It is important to examine the conditions that parameters $\gamma$ and $\eta$ have to satisfy, for the metric $g_{AB}$ to be well-defined. First, we want the area radius $R^2$ to be positive:
\begin{align}\label{eq:paramconst1}
R^2= \frac{\gamma}{\Lambda' \gamma -1}>0  && \Leftrightarrow && \Lambda' > \frac{1}{\gamma}.
\end{align}
Furthermore, $P^2$ also needs to be positive for $x \in (-1,1)$:
\begin{equation}\label{eq:paramconst}
\begin{split}
P^2 =& (1-x^2)\big[1+ \frac{1}{1-\Lambda' \gamma} \frac{1-x^2}{x^2+\eta^2} \big]>0 \\
 & x^2+\eta^2 + \frac{1}{1-\Lambda' \gamma} (1-x^2)>0 \\
\Rightarrow & \hspace{4mm} \gamma<0 \hspace{4mm} \vee \hspace{3mm} \gamma>0\hspace{3mm} \wedge\hspace{3mm} \eta^2> \frac{1}{\Lambda' \gamma - 1}.
\end{split}
\end{equation}
Taking all into consideration, we find that for the cosmological constant greater than zero, $\gamma$ is either negative or positive but then it has to satisfy the following inequalities: $\Lambda' > \frac{1}{\gamma}$ and $\eta^2> \frac{1}{\Lambda' \gamma - 1}$. For negative cosmological constant $\gamma$ has to be negative and satisfy: $\Lambda' > \frac{1}{\gamma}$. These constraints will be used in the following subsection to find a range of values  for the area and the angular momentum. 

\subsection{Area and angular momentum}\label{AJ}
 We now have the tools to calculate the area $A$ and angular momentum $J$ in terms of the rescaled cosmological constant $\Lambda '$ as well as parameters $\gamma$ and $\eta$. Area can be obtained using eq. (\ref{eq:r}): 
\begin{equation}
\begin{split}
A = 4\pi R^2 = 2\pi \frac{\gamma}{\gamma \Lambda' - 1},
\end{split}
\end{equation}
whereas the angular momentum can be found using the imaginary part of the Weyl coefficient $\Psi_2$:
\begin{equation}
\begin{split}
 J &= - \frac{1}{4\pi} \int_{S} \phi {{\rm Im}\Psi_2} {}^{{}^{(2)}}\epsilon \\ 
 &= \frac{1}{8\pi}  \int_0 ^{2\pi} d\varphi \int_{-1}^{1} dx P^2 \partial_x U \partial_x \phi  \\ 
 &= \frac{1}{4(\gamma\Lambda'-1)^2}   \frac{\gamma}{\eta},
\end{split}
\end{equation}
where we have used integration by parts and constrain (\ref{eq:conditionU}). Function $\phi$ is defined up to an additive constant as the generator of the vector field $\Phi = \partial_\varphi$:
\begin{equation}
\phi_{,B} := {\Phi^A} \ {}^{{}^{(2)}}\epsilon_{AB}.
\end{equation}
Our next goal is to find whether the found solutions could be parametrized by area $A$ and angular momentum $J$. First, we notice that:
\begin{align}
A_1 =A_2 \ \ &\Leftrightarrow  \ \gamma_1 = \gamma_2 \ ; \nonumber \\
J_1=J_2 \ \  &\Leftrightarrow   \ \eta_1 = \eta_2 \nonumber \ .
\end{align}
Moreover, considering parameters' constraints (\ref{eq:paramconst}) for the cosmological constant greater than zero we obtain that for negative $\gamma$:
\begin{align}
 A \in \bigg(0, \ \frac{ 2 \pi}{\Lambda '} \bigg) && J \in \big(-\infty, \ 0\big)\ \vee \ \big(0, \ \infty \big),
 \end{align}
 and positive $\gamma$:
 \begin{align} A \in \bigg(\frac{2\pi}{\Lambda '}, \ \infty \bigg)  && \mid J \mid \ \in \Big(0, \ \frac{A}{8\pi} \sqrt{\frac{\Lambda' A}{2\pi} -1} \Big).
 \end{align}
  For negative cosmological constant there is just one case, since $\gamma$ cannot be positive, namely: \begin{align} A \in \big(0, \ \infty \big) && J \in \big(-\infty, \ 0 \big)\ \vee \ \big(0, \ \infty \big).
  \end{align}
It follows, that there is a one-to-one correspondence between area $A$ and angular momentum $J$ and parameters $\eta$ and $\gamma$. Therefore, the specific values of $A$ and $J$ uniquely determine the type D horizon.

\subsection{Embeddability in  the generalized Kerr-(anti) de Sitter spacetimes} \label{KdS}
Consider the $2$-dimensional family of axisymmetric vacuum solutions to Einstein's equations with the cosmological constant
\begin{equation}
\begin{split}\label{KdSmetric}
ds^2 = - \frac{\Delta_r}{\chi^2 \rho^2}(dt-a \sin^2\theta d\varphi)^2 + \frac{\Delta_\theta \sin^2 \theta}{\chi^2 \rho^2}[adt-(r^2+a^2)d\varphi]^2 + \rho^2 \bigg( \frac{dr^2}{\Delta_r} + \frac{d\theta^2}{\Delta_\theta} \bigg),
\end{split}
\end{equation}
where the parameters are: the mass $M$ and a parameter $a$ corresponding to the angular momentum, and 
\begin{align}
\rho^2 &= r^2 + a^2\cos^2\theta,\\ 
\Delta_\theta &= 1 + \frac{1}{3}\Lambda a^2 \cos^2\theta, \\
\chi&=1+\frac{1}{3}\Lambda a^2,\\
\Delta_r &= \Big(r^2+a^2\Big)\Big(1-\frac{1}{3} \Lambda r^2\Big) - 2Mr.
\end{align}
All the metric tensors in this family are of the Petrov type D.  Every solution $r_{_0}$ to the equation 
\begin{equation}\label{deltaR}
 \Big(r_{_0}^2+a^2\Big)\Big(1-\frac{1}{3} \Lambda r_{_0}^2\Big) - 2Mr_{_0} = 0,
\end{equation}
defines a Killing horizon in a suitably extended  spacetime. For $\Lambda>0$ some of the roots may be negative, which means that the coordinate $r$ ranges negative numbers.  
Each of those horizons is of the type D.
Hence, as long as it is non-extremal, it is one of our solutions derived above.  Given our solution $P,U$ (\ref{eq:p2},\ref{eq:U})
of the area radius $R$, we find a corresponding Killing horizon in (\ref{KdSmetric}) by matching the parameters $M$ and $a$. First, 
we compare the metric tensors on  $2$-dimensional cross-sections    of the horizons, namely
\begin{equation}\label{metric}
R^2\bigg(\frac{1}{P^2} dx^2 + P^2d\varphi^2 \bigg)= \frac{\rho^2}{\Delta_\theta} d\theta^2 + \frac{\Delta_\theta (r_{_0}^2 + a^2)^2}{\chi^2 \rho^2} \sin^2 \theta d\varphi^2 .
\end{equation}
Given the left hand side, we consider it an equation for the right hand side. Calculating the total areas we find
\begin{equation}
 R^2=(r_{_0}^2+a^2)/\chi.
 \end{equation} 
 Next, we compare the corresponding coefficients:
\begin{equation}\label{eq:comp1}
\begin{split}
 \frac{\Delta_\theta (r_{_0}^2 + a^2)^2}{\chi^2 \rho^2} \sin^2 \theta & =  R^2 P^2 \\
 \frac{\Delta_\theta}{\rho^2} R^2 \sin^2 \theta &= P^2,
  \end{split}
 \end{equation}
 
 Finally, we use the above expression for $P^2$ to obtain the function $x(\theta)$:
 \begin{equation}
\begin{split}
\frac{R^2}{P^2} dx^2 &=  \frac{\rho^2}{\Delta_\theta} d\theta^2 \\
 \frac{\rho^2}{\Delta_\theta \sin^2 \theta} dx^2 &=  \frac{\rho^2}{\Delta_\theta} d\theta^2  \\
 x&=-\cos\theta.
 \nonumber
  \end{split}
 \end{equation}
 Eq. (\ref{eq:comp1}) written explicitly in terms of $x$ reads:
 \begin{equation}
\begin{split}
\frac{1-x^2}{1-\gamma \Lambda/6} \bigg( \frac{(1-\gamma \Lambda/6)(x^2 + \eta^2) + (1-x^2)}{x^2 +\eta^2} \bigg)&=\frac{(1-x^2)(1+\frac{1}{3}\Lambda a^2 x^2)R^2}{r_{_0}^2+a^2 x^2} .
 \end{split}
 \end{equation}
 Comparing coefficients of different powers of $x$ yields that $\eta^2 = r_{_0}^2/a^2$. Therefore, with such $\eta$ the resulting form of $P^2$ (\ref{eq:p2}) agrees with (\ref{metric}). Parameters $\gamma$ and $\eta $ expressed in terms of $a$ and $r_{_0}$ as well as area radius $R$ and angular momentum $J$ take the following form:
\begin{align}
\gamma=& \frac{6(a^2+r_{_0}^2)}{r_{_0}^2 \Lambda -3}= \frac{2R^2}{\frac{1}{3}\Lambda R^2 -1}, \\
\eta^2 &= \frac{r_{_0}^2}{a^2} =\frac{R^4}{4J^2}(1-\frac{1}{3}\Lambda R^2)^2.
\end{align}
Analogously, expressing KdS (KadS) parameters $r_{_0}$ and $a$ in terms of $\gamma$ and $\eta$ yields:
\begin{align}
\label{r}
r_{_0}^2 &= \frac{3\gamma}{\Lambda \gamma - 6(\frac{1}{\eta^2} +1)}, \\
\label{a}
a^2 &= 
 \frac{3\gamma}{\Lambda \gamma \eta^2 - 6(\eta^2 +1)}.
\end{align}   
Now we check whether different pairs of parameters $\eta$ and $\gamma$ could lead to the same $r_{_0}$ and $a$ on the KdS (KadS) side:
\begin{equation}
\begin{split}
r_{{_0} 1}^2 = r_{{_0} 2}^2 &\hspace{4mm}  \Leftrightarrow  \hspace{4mm} \frac{3\gamma_1}{\Lambda \gamma_1 - 6(\frac{1}{\eta_1^2}+1)}=\frac{3\gamma_2}{\Lambda \gamma_2 - 6(\frac{1}{\eta_2^2}+1)}  \hspace{4mm}  \Leftrightarrow  \hspace{4mm} \gamma_2= \gamma_1 \frac{\eta_1^2(1+\eta_2^2)}{\eta_2^2(1+\eta_1^2)},\\
a_1^2=a_2^2  &\hspace{4mm}  \Leftrightarrow  \hspace{4mm} \frac{3\gamma_1}{\Lambda \gamma_1 \eta_1^2 - 6(\eta_1^2 +1)}=\frac{3\gamma_2}{\Lambda \gamma_2 \eta_2^2 - 6(\eta_2^2 +1)} \\
&\hspace{4mm}  \Leftrightarrow  \hspace{4mm} \gamma_1 (1+\eta_2^2)(\eta_2^2-\eta_1^2)(\frac{\Lambda \gamma_1 \eta_1^2}{1+\eta_1^2} -6) = 0\\
&\hspace{4mm}  \Leftrightarrow  \hspace{4mm} \eta_1^2=\eta_2^2 \hspace{4mm}  \vee \hspace{4mm}  \eta_1^2 = \frac{1}{\Lambda' \gamma_1 - 1}.
\nonumber
\end{split}
\end{equation}
Therefore, there are two possibilities for such a scenario, first when $\eta_1^2=\frac{1}{\Lambda'\gamma_1 - 1}$, which however does not satisfy constraint (\ref{eq:paramconst}). The second case is $\eta_1=-\eta_2$, where the sign difference comes from the opposite direction of rotation ($a_1=-a_2$). 
We have also found that parameters $r_{_0}^2$ and $a^2$ can take arbitrary positive values. Consequently, there is a one-to-one correspondence between the KdS (KadS) horizon parameters and parameters characterizing the considered type D horizon. It follows that the obtained solutions are embeddable in the Kerr-(anti) de Sitter metric provided that the corresponding horizon in the  KdS (KadS) spacetime  is non-extremal.
For every value of parameters $\gamma$ and $\eta$ satisfying constraints (\ref{eq:paramconst1}) and (\ref{eq:paramconst}), parameter $M$ in (\ref{KdSmetric}) can be obtained using equation (\ref{deltaR}) and reads:
\begin{equation}\label{m}
M=\frac{(1+\eta^2)^2}{2\sqrt{2} \eta^2}\sqrt{\frac{-\gamma \eta^2}{\bigg[ 1+\eta^2 \big(1-\frac{\Lambda}{6} \gamma \big) \bigg]^3}}.
\end{equation}
Although $r$ ranges negative numbers,  using the transformation $r_{_0}\mapsto -r_{_0}$ and $M\mapsto -M$ we may keep $M$ or $r_{_{0}}$ positive. 
The Killing horizon corresponding to $r= r_{_0}$ in the spacetime of the metric tensor (\ref{KdSmetric})  defined by the established
values of $a$ and $M$ is one of our type D horizons provided it is non-extremal. However, in this case, the condition (\ref{extrem})
is more difficult to verify. The extremality can be investigated by analyzing the dependence of the roots of the coefficient $\Delta_r$
on the parameters $M,a$ and $\Lambda$. The problem  is soluble, however the solution does not give us explicit conditions on those parameters.   In the extremal case, the corresponding  Killing horizon in (\ref{KdSmetric}) is not our non-extremal horizon. In that
case our non-extremal type D horizon is embeddable in the near extremal horizon limit spacetime \cite{Horowitz,NHG}. 

 \subsection{No hair theorem for the Petrov type D axisymmetric isolated horizons}
Concluding our work we formulate a no hair theorem for the Petrov type D axisymmetric null surfaces of topologically spherical sections:
\medskip
 
\noindent {\bf Theorem} (No-hair) {\it On a topological  $2$-sphere $S$, every axisymmetric solution $g_{AB}$ and ${\Omega}_{AB}$ to the Petrov 
type D equation (\ref{eq:typeD1}) with a cosmological constant $\Lambda$  is uniquely determined by a pair of numbers: the area $A$, and the angular momentum $J$.  The range of the pairs $(A,J)$ corresponding to the solutions depends on the   cosmological
constant as follows: 
\begin{itemize}
\item for $\Lambda > 0,$ $J \in (-\infty,\infty)$ for $A \in \big(0, \frac{ 12 \pi}{\Lambda} \big)$ and $\mid J \mid \ \in \Big[0, \frac{A}{8\pi} \sqrt{\frac{\Lambda A}{12\pi} -1} \Big)$ for $A \in \big(\frac{12\pi}{\Lambda}, \infty \big)$;
\item for $\Lambda\le 0$,   $J \in (-\infty,\infty)$ and $A \in (0, \infty)$.
\end{itemize}
Every solution defines a type D isolated horizon whose intrinsic geometry $(g_{ab}, \nabla_a)$ coincides with the intrinsic geometry of 
a non-extremal Killing horizon contained  in one of the following (locally defined) spacetimes: 
(i) the generalized Kerr  - (anti) de Sitter spacetime (\ref{KdSmetric}),  (ii) the generalized  Schwarzschild - (anti) de Sitter spacetime 
(\ref{SdS}), (iii)  the near horizon limit spacetime near an extremal horizon contained either in the generalized Kerr  - (anti) de Sitter 
spacetime or in  the generalized  Schwarzschild - (anti) de Sitter spacetime (see \cite{NHG,LSW}).}    
\medskip

This is a generalization of a similar result derived in the case of the vanishing cosmological constant \cite{LPkerr}. 
 
\section{Summary}   We have considered  non-extremal isolated horizons embeddable in 4-dimensional spacetimes satisfying the vacuum Einstein equations with a non-vanishing cosmological constant. The Weyl tensor at the horizon was assumed to be of the Petrov type D and the stationarity to the second order assumption made it determined by the intrinsic geometry of the horizon.
We have found the general solution to  the Petrov type D equation \cite{DLP1} on an unknown metric $g_{AB}$ and the exact 2-form 
$\Omega$   defined on a topological 2-sphere  in the axisymmetric case.   The components of IH geometry have been expressed in Newman-Penrose null frame. We have introduced the coordinate system adapted to the axial symmetry and we have found that the resulting family of solutions is 2-dimensional and can be uniquely parametrized by the area and the angular momentum which might be regarded as a no hair theorem for the Petrov type D axisymmetric IHs. The comparison with the generalized  Kerr-(anti) de Sitter horizon has been performed. We found the relation between parameters $a$ and $r_{_0}$ from the horizon slice of generalized  KdS (KadS) metric written in Boyer-Lindquist coordinates and the parameters $\gamma$ and $\eta$  defining our solutions. 
Most of the obtained IHs are embeddable in the generalized Schwarzschild / Kerr-(anti) de Sitter  metrics. The exceptions are those admitting another null symmetry that is extremal. In this case they are embeddable in the near extremal horizon spacetimes \cite{Horowitz} constructed from the generalized  Schwarzschild / Kerr-(anti) de Sitter metrics. This work has been a generalization of the previous result for the axisymmetric type D isolated horizon in the cosmological constant free spacetime \cite{LPkerr}. Our solutions to the Petrov type D equation provide a local characteristic for the spacetime, which distinguishes the Kerr-(anti) de Sitter solutions. For example it may be applied to the black hole holograph construction of spacetimes about Killing horizons that uses as the starting point exactly the same data $g_{AB}$ and $\Omega_{AB}$ on a $2$-dimensional surface \cite{R1, R2}.
  
  \vspace{1cm}

\noindent{\bf Acknowledgements:}
This work was partially supported by the Polish National Science Centre grant No. 2015/17/B/ST2/02871.

\section{Appendix}
\subsection{Abstract notation}\label{notation}
W have used the same abstract index notation \cite{Wald}, as in the accompanying paper \cite{DLP1}:
\begin{itemize}
\item  Indices of the spacetime tensors  are denoted by
lower Greek letters: $ \alpha, \beta,\gamma, ... \ = 1,2,3,4$
\item  Tensors defined in $3$-dimensional  space $H$  carry indexes denoted by lower Latin letters: $a, b, c, ...\ = 1,2,3$
\item  Capital Latin letters $A, B, C, ... \ = 1,2$ are used as the indexes of tensors defined in $S$.  
\end{itemize}

\subsection{Complete description in the Newman-Penrose Null Frame}
The complex null vector frame $(e_1,\dots,e_4)=(m,\bar{m},n,\ell)$ 
(where $n,l$ are real vectors and $m$ is complex) of a four-dimensional
spacetime consists the 
Newman-Penrose null tetrad if the following scalar products
\begin{subequations}\label{eq:NP-prod}\begin{align}
  g^{\mu\nu}m^{\mu}\bar{m}^{\nu} &= 1  &
  g^{\mu\nu}n^{\mu}\ell^{\nu} &= -1  \tag{\ref{eq:NP-prod}}
\end{align}\end{subequations}
are the only non-vanishing products of the frame components. The dual
frame corresponding to the tetrad is denoted as
$(e^1,\dots,e^4)$. In terms of this coframe components the metric
tensor takes the form:
\begin{equation}\label{eq:NP-metric}
  g_{\mu\nu}\ =\ e^1{}_{\mu}e^2{}_{\nu} + e^2{}_{\mu}e^1{}_{\nu}
    - e^3{}_{\mu}e^4{}_{\nu} - e^4{}_{\mu}e^3{}_{\nu}
\end{equation}

The torsion-free spacetime connection is determined by the 1-forms defined 
as follows
\begin{subequations}\label{eq:NP-con-def}\begin{align}
  \Gamma_{\alpha\beta}\ &=\ - \Gamma_{\beta\alpha} &
  \\ e^{\alpha} + \Gamma^{\alpha}{}_{\beta}\wedge e^{\beta}\ &=\ 0
    \tag{\ref{eq:NP-con-def}}
\end{align}\end{subequations}
which can be decomposed onto the following complex coefficients
\begin{subequations}\label{eq:NP-con}\begin{align}
  - \Gamma_{14} &= \sigma e^1 + \rho e^2 + \tau e^3 + \varkappa e^4 \\
    \Gamma_{23} &= \mu e^1 + \lambda e^2 + \nu e^3 + \pi e^4 \\
  - \frac{1}{2}(\Gamma_{12} + \Gamma_{34})
   &= \beta e^1 + \alpha e^2 + \gamma e^3 + \epsilon e^4
\end{align}\end{subequations}
called the spin coefficients. 

We can choose $n$ such that:
\begin{equation}
\nabla_n e^\mu = 0,
\end{equation}
and from which it follows that \cite{LPnei}: 
\begin{equation}
\tau = \nu = \mu- \bar \mu = \pi - (\alpha + \bar \beta) = 0.
\end{equation}

\bibliography{drlp-ih}{}
\bibliographystyle{apsrev4-1}
\end{document}